\documentclass[letterpaper,twocolumn,10pt]{article}
\usepackage{usenix2019_v3}
\usepackage{floatrow}
\usepackage{tikz}
\usepackage{biblatex}
\AtBeginBibliography{\small}
\usepackage{amsmath}
\usepackage{censor}
\usepackage{dingbat}
\usepackage{tikz}
\usepackage{subfigure}
\usepackage{pifont}
\newcommand{\cmark}{\ding{51}}%
\newcommand{\xmark}{\ding{55}}%

\usepackage{enumitem}

\usepackage{booktabs}
\usepackage{fontenc}
\usepackage{pifont}

\usepackage{filecontents}
\usepackage{cleveref}
\usepackage{cellspace}
\usepackage{url}
\begin{document}

\date{}

\title{\Large \bf  Account Verification on Social Media: User Perceptions and Paid Enrollment}

\author{
 {\rm Madelyne Xiao}\\
 Princeton University
 \and
 {\rm Mona Wang}\\
 Princeton University
 \and
 {\rm Anunay Kulshrestha}\\
 Princeton University
 \and
 {\rm Jonathan Mayer}\\
 Princeton University
} 

\maketitle

\begin{abstract}
We investigate how users perceive social media account verification, how those perceptions compare to platform practices, and what happens when a gap emerges. We use recent changes in Twitter's verification process as a natural experiment, where the meaning and types of verification indicators rapidly and significantly shift. The project consists of two components: a user survey and a measurement of verified Twitter accounts.

In the survey study, we ask a demographically representative sample of U.S. respondents ($n=299$) about social media account verification requirements both in general and for particular platforms. We also ask about experiences with online information sources and digital literacy. More than half of respondents misunderstand Twitter's criteria for blue check account verification, and over 80\% of respondents misunderstand Twitter's new gold and gray check verification indicators. Our analysis of survey responses suggests that people who are older or have lower digital literacy may be modestly more likely to misunderstand Twitter verification.

In the measurement study, we randomly sample 15 million English language tweets from October 2022. We obtain account verification status for the associated accounts in November 2022, just before Twitter's verification changes, and we collect verification status again in January 2022. The resulting longitudinal dataset of 2.85 million accounts enables us to characterize the accounts that gained and lost verification following Twitter's changes. We find that accounts posting conservative political content, exhibiting positive views about Elon Musk, and promoting cryptocurrencies disproportionately obtain blue check verification after Twitter's changes.

We close by offering recommendations for improving account verification indicators and processes.
\end{abstract}

\section{Introduction}
Soon after acquiring Twitter in October 2022, Elon Musk rebooted the platform's account verification process~\cite{mehta-musk-2022}. The headline change was a new subscription model for blue check verification, where U.S. users could pay $\$8$ per month to enroll. Previously, Twitter only awarded blue checks to accounts that were ``active,'' ``notable,'' and ``authentic''~\cite{legacy-verif}. Demonstrating authenticity involved proof of identity, such as a photo of a driver's license. Under the new regime, Twitter relaxed these requirements, including by replacing ``authentic'' with ``secure'' (older than 90 days\footnote{Twitter subsequently relaxed this verification requirement further, to an account age of 30 days. Twitter's continually shifting verification requirements posed a challenge for this research, which we discuss as a limitation.} plus a confirmed phone number) and ``non-deceptive'' (i.e., not impersonating a person or organization, not spam, and no recent name or photo changes)~\cite{twitter_verified_accounts}. Crucially, the new verification process for blue checks did not---and still does not---require affirmative proof of identity.

Predictably, some users took advantage of paid verification to spoof high-profile Twitter accounts. In a particularly notable instance, a fake Eli Lilly account (\texttt{@EliLillyandCo}) obtained a blue verification check and tweeted that ``insulin is now free''~\cite{fake_lilly_tweet_2022}. Patients and healthcare providers responded to the announcement, which quickly went viral, with hope and confusion. Twitter finally took down the account about 8 hours later, after Eli Lilly's market cap had dropped by billions of dollars on unusually high trading volume.\footnote{The tweet was partly satirical, highlighting the high cost of insulin. The causal relationship between the tweet, Eli Lilly stock, and healthcare sector market movement is difficult to ascertain and beyond the scope of this work.}

The potential economic effects  of misleading verification are deeply concerning. Harms to public trust, especially in crisis situations where official communications are critical, are perhaps even more alarming and less quantifiable. The popularity of social media and proliferation of bots and fake accounts suggest that verification has never been more vital.

\begin{table*}[ht!]
\footnotesize
\begin{tabular}{llrclccc}
\toprule
Platform & Program & Launch & Icon & Eligibility Criteria & Open & Requestable & Pay\\
\midrule
Twitter & Legacy Verification & 2009 & \includegraphics[height=\fontdimen6\textfont2]{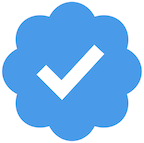} & ``Active,'' ``Notable,'' and ``Authentic'' & \xmark & \cmark & \xmark\phantom{${}^{*}$}\\
Twitter & Blue & 2022 & \includegraphics[height=\fontdimen6\textfont2]{images/check_tw_blue.png} & ``Complete,'' ``Active,'' ``Secure,'' and ``Non-deceptive'' & \cmark & \cmark & \cmark\phantom{${}^{*}$}\\
Twitter & Verified Organizations & 2022 & \includegraphics[height=\fontdimen6\textfont2]{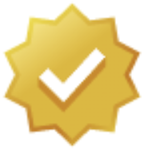} & Organizational account, additional criteria ambiguous & \cmark & \cmark & \xmark${}^{*}$\\
Twitter & Government & 2022 & \includegraphics[height=\fontdimen6\textfont2]{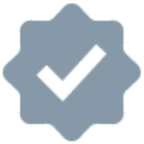} & Specified account types, additional criteria ambiguous & \cmark & \cmark & \xmark\phantom{${}^{*}$}\\
\midrule
Facebook & Verified Pages and Profiles & 2013 & \includegraphics[height=\fontdimen6\textfont2]{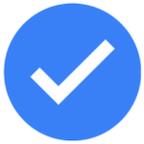} & ``Notable,'' ``Unique,'' and ``Authentic''${}^{*}$ & \cmark & \cmark & \xmark${}^{*}$\\
Instagram & Verified Badges & 2014 & \includegraphics[height=\fontdimen6\textfont2]{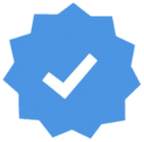} & ``Notable,'' ``Unique,'' and ``Authentic''${}^{*}$ & \cmark & \cmark & \xmark${}^{*}$\\
\midrule
Snapchat & Snap Stars & 2019 & \includegraphics[height=\fontdimen6\textfont2]{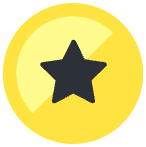} & ``Engagement,'' ``Public Audience,'' ``Authenticity,'' & \cmark & \xmark & \xmark\phantom{${}^{*}$}\\
 & & & &  ``Notability,'' and ``Quality'' & & \\
Snapchat & Public Profiles for Businesses & 2021 & \includegraphics[height=\fontdimen6\textfont2]{images/check_snap.png} & ``Authentic'' and ``Notable'' & \cmark & \cmark & \xmark\phantom{${}^{*}$}\\
\midrule
TikTok & Verified Accounts & 2019 & \includegraphics[height=\fontdimen6\textfont2]{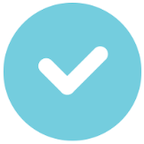} & ``Active,'' ``Authentic,'' ``Complete,'' ``Notable,'' and & \cmark & \cmark & \xmark\phantom{${}^{*}$}\\
 & & & & ``Secure'' \\
\midrule
YouTube & Channel Verification & 2022 & \includegraphics[height=\fontdimen6\textfont2]{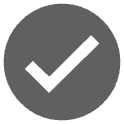} & ``Authentic,'' ``Complete,'' and 100,000 subscribers & \cmark & \cmark & \xmark\phantom{${}^{*}$}\\

\bottomrule
\end{tabular}\\

\caption{Account verification programs at major social media platforms as of January 2023, when we conducted the survey study~\cite{tw-gold-verif-2023, twitter_verified_accounts, fb-verif-2023, tt-verif-2023, yt-verif-2023, ig-verif-2023}. The ``Open'' column indicates whether a program was enrolling new accounts, ``Requestable'' indicates whether accounts could ever request to be verified through a standardized process, and ``Pay'' indicates whether a subscription could be a component of enrollment. An asterisk denotes an aspect of a verification program that changed between the survey and publication: Twitter Verified Organizations became available for pay, and Meta both eliminated its ``Notable'' and ``Unique'' requirements and made verification available for pay.}

\label{table:verif_marks}
\end{table*}

Twitter pioneered social media account verification in 2009, after determining that after-the-fact takedowns were insufficient to address imposter accounts~\cite{stone-tweet-2009}. In the following years, major social media platforms have generally followed Twitter's model, offering account verification with similar terminology and check mark iconography~\cite{fb-verified-2013,chang-ig-2014,tiktok-verif}. Table~\ref{table:verif_marks} summarizes account verification programs offered by major social media platforms at the time of this research.

The primary purpose of account verification---a point of consistency over time and across platforms, with the sole exception of Twitter's recent changes---has been to address account impersonation by affirmatively and proactively confirming an account owner's identity. But verification quickly accumulated other meanings, including as a signifier of account importance and the credibility of content~\cite{kabakus-tw-users-2019, vaidya-2019-verified-credibility}.

The ongoing fallout from Twitter's verification changes, differences in social media platform practices, and varied possible meanings of account verification underscore the need for in-depth study both of how users perceive verification and how behaviors shift in response to verification changes. Our project is also situated in extensive literature about trust indicators and perceptions of security, work that has demonstrated how perceptions can differ from semantics and identified best practices for effective indicators. Section~\ref{sec:related_work} provides an overview of related work.

Against this backdrop, we make two complementary contributions, which we motivate and outline in Section~\ref{sec:motivation}.

First, we conduct a U.S. demographically representative survey ($n = 299$) to understand how people perceive social media account verification both in general and for specific platforms. The survey also asks about online information sources and digital literacy. The results demonstrate a significant mismatch between perceptions and reality: more than half of respondents misunderstand Twitter's blue check verification policies to still require proof of identity, and over 80\% of respondents misunderstand Twitter's gold and gray check indicators. Correlation analysis of survey responses suggests that people who are older or have lower digital literacy may be modestly more likely to misunderstand Twitter verification. Section~\ref{sec:survey} presents the survey design and results.

Our second main contribution is a Twitter measurement study, in which we randomly sample 15 million English language tweets from October 2022. We obtain verification status for the posting accounts in November 2022, just prior to Twitter changing its verification practices, and we collect verification status again in January 2022. The resulting longitudinal dataset of 2.85 million accounts allows us to characterize trends in account verification after Twitter's changes. We find that Twitter accounts posting conservative perspectives, sharing favorable views on Elon Musk, and boosting cryptocurrencies more commonly gain blue check verification after the changes. The cryptocurrency results in particular suggest ongoing strategic exploitation of how people misunderstand verification. Section~\ref{sec:measurement} presents the measurement study.

Taken together, the findings from these studies highlight shortcomings of Twitter ``Blue'' paid verification  (Section~\ref{sec:improvements}). The results suggest that Twitter's changes may run afoul of consumer protection law---in particular, Section 5 of the Federal Trade Commission Act and Twitter's consent order with the FTC from May 2022 \cite{ftc-sec5-2008, ftc-consent-2022}. Our work also informs general best practices for social media account verification.

\begin{figure*}[!htb]
\begin{subfigure}
\noindent\makebox[\textwidth]{\includegraphics[scale=0.48]{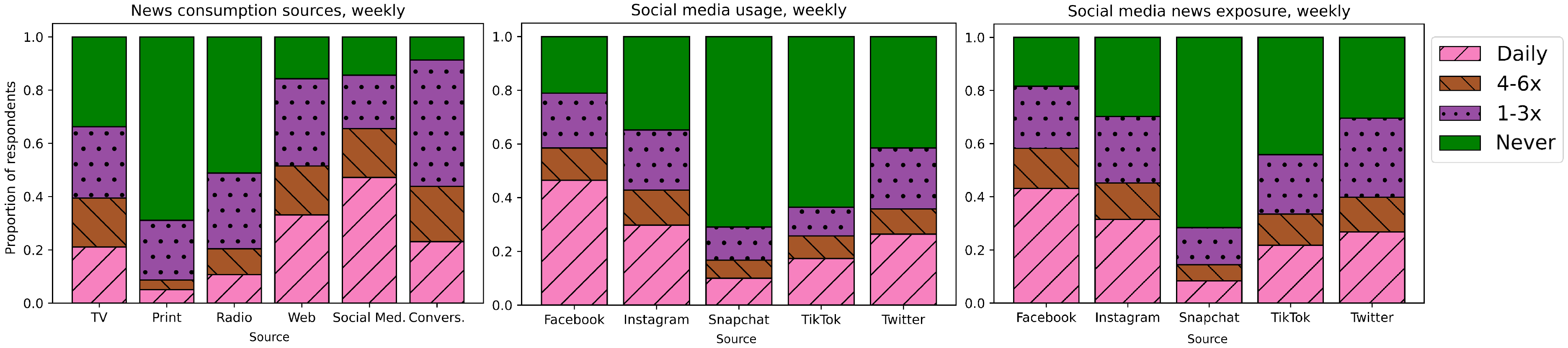}}
\label{fig:usage}
\caption{News consumption and social media usage. Survey respondents described the frequency with which they use personal social media accounts, engage with news stories on social media, and consume news stories via different channels.}
\end{subfigure}
\end{figure*}

\section{Related Work} 
\label{sec:related_work}

This research builds on limited prior work about social media account verification and extensive scholarship about trust indicators and security perceptions.

\paragraph{Social media account verification.} 

There is surprisingly little literature on perceptions and effects of social media account verification, despite the prevalence of these indicators and their potentially confusing nature. In a 2019 publication that is close to our work, Vaidya et al. studied the effects of the Twitter verified badge on an account's tweet credibility~\cite{vaidya-2019-verified-credibility}. The researchers evaluated the effect of verification indicators on perceived credibility and self-reported sharing intentions, finding in a pair of surveys that people do not interpret verification as an indicator of credibility and do not self-report greater likelihood of sharing content from verified accounts. Similarly, in a study from 2019, Edgerly and Vraga found that verification marks on Twitter did not significantly influence users' evaluations of the credibility of a news-related tweet or its source material \cite{vraga_2019}. Our survey study, by contrast, centers on the requirements for an account to obtain verified status and the natural experiment of Twitter's verification changes. We also explore differences in verification across platforms, demographic and digital literacy correlates of perceptions, user experiences with account verification, and user preferences for verification semantics.

\paragraph{Trust indicators and warnings for website identity.} Our study is adjacent to a large body of literature on trust indicators and warnings as a defense against malware, person-in-the-middle attacks, phishing, social engineering attacks, and other online risks. In general, such attacks exploit user misperceptions about the trustworthiness of content, services, and applications. Trust indicators and warnings address these risks by providing users with relevant information, typically an authenticity check based on digital signatures for trust indicators and other security checks for warnings.

A social media account verification mark is, fundamentally, a type of trust indicator. Verification marks are also conceptually analogous to certain security warnings. We expand on these analogies when providing a threat model in Section~\ref{sec:threat_model}.

Much prior work has examined ways to effectively communicate website identity, to prevent website impersonation, and to prevent other phishing or security risks to end users~\cite{browsers-against-phishing-2008}. For instance, modern web browsers use security indicators and warnings to warn users of potential malware, phishing, or person-in-the-middle attacks. Such interventions have been extensively evaluated both in the lab and in the real world~\cite{egelman2008, sunshine2009, krol2012, akhawe2013}. Researchers have also found that in practice, users conflate encryption and authentication in their mental models of security indicators in the website security context~\cite{krombholz-2019-https-models}.  

\paragraph{User perception of online security.} User perception of security has been studied in a wide variety of online contexts. For example, Dechand et al. study user perception of end-to-end encryption on WhatsApp and find that users largely do not trust it~\cite{dechand2019}. In a systematic review of user studies of multi-factor authentication, Das et al. consistently find low adoption~\cite{das2019}. Ur et al. investigate whether user perception of password security match reality and find significant differences across users' understanding of possible attacks~\cite{ur2016}.

Perceptions of social media users have also been studied. 
However, work in this area has focused on perceptions of social media website quality~\cite{ellahi2013}, 
level of control over information shared~\cite{hajli2016, saridakis2016}, and protection from abuse and harrassment~\cite{redmiles2019}.
Usability of security features on social media platforms has only been analyzed in the context of security notices. Benson et al. find that users disclose more information in their presence~\cite{benson2015}.\\

The methods in this project are inspired by both areas of related work. We examine social media account verification with complementary methods, similar to prior work that uses both surveys and implementation measurements. The survey illuminates perceptions both in general and in the natural experiment of Twitter’s changes, and the measurement quantifies behavior following the changes.

\section{Motivation and Research Questions}
\label{sec:motivation}

Social media has become an important channel for communicating with friends, following celebrities and influencers, and keeping up with news. According to a 2022 Pew survey, more than 70\% of American adults received their news from social media platforms; about 27\% of American adults said that they regularly receive their news via Twitter \cite{pew-sm-report-2022}. These results are borne out by our own survey: among the 299 respondents represented in our survey, about 50\% said that they receive their news daily via social media, and about 30\% see news stories via Twitter on a weekly basis (see Figure~\ref{fig:usage}).

As previous work in information security research has demonstrated, attentive user interface design has been important to reducing security risks due to impersonation and phishing attacks~\cite{browsers-against-phishing-2008, egelman2008}. In particular, these types of social engineering attacks thrive when there are large discrepancies between the intended meaning of a passive user interface indicator and user perception of such an indicator. In this case, Twitter provides an important case study due to the sudden shift in the semantic meaning of the blue check mark indicator. In addition, very few other social media or messaging platforms use check mark indicators to indicate subscription status, though some, such as GitHub and Signal, use other indicators for financial contributors to projects~\cite{signal_donor_badge, github_sponsors}.

To this end, we believe that our work addresses an urgent need for standardization of verification protocols across social media platforms. En route to producing unified criteria, it is necessary to identify shortfalls in existing verification processes. We conduct two complementary studies to understand the gap between user perception and the reality of social media verification indicators on Twitter.

\subsection{Survey Study Research Questions}
The first study aims to understand current user perceptions of social media verification indicators via a demographically representative survey. The following research questions guided our examination of how people understand verification requirements and processes across social media platforms:

\begin{itemize}
    \item[\textcolor{blue}{RQ1.}] What kinds of assurances do social media users believe existing verification marks provide? 

    \item[\textcolor{blue}{RQ2.}] Is there a mismatch between expected and actual assurances provided by verification marks?
    
    \item[\textcolor{blue}{RQ3.}] Do respondents perceive Twitter's current practices---including confirmed receipt of a text or call---to be sufficient for account  verification?

    \item[\textcolor{blue}{RQ4.}] How do perceptions of social media verification marks interact with demographics?  

\end{itemize}

\subsection{Measurement Study Research Questions}
\label{sec:data-driven-analysis-rq}
The second study then aims to understand trends in Twitter verification by collecting, annotating, and analyzing a dataset of accounts on the platform. The following research questions guided our examination of verified Twitter accounts:

\begin{itemize}
\item[\textcolor{blue}{RQ5.}] What is the composition of accounts with blue check verification before and after Twitter's changes in November 2022?
\item[\textcolor{blue}{RQ6.}] What types of content are published by accounts with blue check verification before and after Twitter's changes?

\end{itemize}

For each study, we describe methods, limitations,  results, and implications. We conclude with recommendations for improving current social media verification processes.

\subsection{Threat Model}
\label{sec:threat_model}

Before turning to our studies, we briefly offer a formal threat model for social media account verification. We derive this threat model from social media platform descriptions of account verification processes and goals, with the exception of Twitter's paid Blue verification service.

Account verification addresses two specific and related risks. First, an account could misrepresent the person or organization that controls the account. This risk parallels domain impersonation risks, and verification marks are conceptually analogous to TLS/SSL indicators. Platform verification parallels certificate authority validation: just as individual, organization, and extended validation aim to prevent impersonating a person or organization by verifying domain control, social media account verification prevents impersonation by confirming account control.

The second risk that social media verification addresses is an account impersonating another account with a similar handle or name. This risk parallels phishing, homographs, typosquatting, and similar attacks that depend on using a similar domain name or email address to mislead users. Social media account verification is somewhat analogous to web browser and online service warnings that mitigate these risks. 

Importantly, account compromise is \textit{not} in the threat model for social media account verification. A verified account may be compromised by a malicious party, just as a website with an authentic certificate and legitimate ownership of a domain may be compromised.

\section{Survey Study: Perceptions of Verification}
\label{sec:survey}

We recruited a sample of 300 U.S.-based adult respondents via Prolific, an online survey platform \cite{prolific}. The sample was representative of the U.S. population, as estimated by the Census Bureau, in terms of gender, age, and race.

We recorded respondent demographic, location, and device information. In addition to the survey questions described below, we included three attention-check questions. We compensated survey respondents \$3.75, based on a \$15 per hour wage prorated for a 15-minute expected completion time.\footnote{We calibrated the expected completion time with pilot studies.}

We designed a 37-question survey with a blend of multiple-choice, matrix, and free-response questions. The survey instrument is available as supplementary information. Survey questions addressed the following topics: digital literacy, social media use, perceived requirements for account verification on different platforms, and the evolution of these perceptions over time (specific to Twitter). Details about the survey design and analysis for each research question, corresponding to those listed in the previous section, are below: 

\textcolor{blue}{RQ1.} Respondents selected, from a list of nine options, all criteria that they believe apply to verified accounts on Facebook, Twitter, and TikTok. We developed these options to fairly capture both current verification policies and common misconceptions associated with verification marks. Analysis also included inductive coding of free response answers to a question asking what verification means.

\textcolor{blue}{RQ2.} Using responses to the multiple choice prompt we developed for the previous RQ, we analyze respondent perceptions of \textit{identity} and \textit{payment} as criteria for verification across three different social media platforms and in general. Respectively, these two verification criteria represent the implicit objective of account verification on social media and the most recent update to Twitter's verification policy. 

\textcolor{blue}{RQ3.} We analyzed responses to survey questions about the perceived (in)sufficiency of Twitter Blue's verification criteria. Without naming Twitter, we presented survey respondents with the criteria for obtaining Blue paid verification. Respondents also described their perceptions of Twitter's taxonomy of identification marks, which includes different colored check marks for different types of organizations: gray for government, gold for organizations, and blue for both subscription accounts and ``legacy verified'' accounts (see~\Cref{table:verif_marks}). 

\textcolor{blue}{RQ4.} We conducted bivariate correlation analysis and multivariate regression analysis to understand possible interaction between demographic attributes and perceptions of social media verification. Our analysis focused on digital literacy and age, because anonymous data was readily obtainable and because prior work suggests that these are important factors in security perceptions and online behavior.

We randomized survey questions as necessary to reduce potential biases resulting from question order.

We conducted two pilot studies and a full-scale study via the Prolific online surveying platform~\cite{prolific}. The first pilot study was conducted in December 2022, with 15 respondents, and the second pilot study was conducted in January 2023, with 25 respondents. The full-scale study was conducted in January 2023, with 300 respondents. Below, we describe findings from the main study, organized by relevance to our RQs. Our pilot results are available in~\Cref{sec:pilot}.

\subsection{Ethical Considerations}
This study constitutes human subjects research. Before carrying out the study, we applied for and received approval from the Princeton University Institutional Review Board. We minimized the data that we obtained when carrying out the study, and we did not collect participant names or contact information (except Prolific IDs). All survey data was stored on encrypted and access controlled devices.

\subsection{Results and Discussion}

From our initial dataset of 300 survey responses, we removed any low-quality responses---cases in which the respondent produced egregiously wrong answers to attention check or news-awareness questions. We removed one such response, for a final sample set of 299 responses.\footnote{In the response that we excluded, the participant incorrectly identified the current U.S. President as Bill Clinton.}

We begin our analysis with a review of our inductive coding of respondents' free responses to a question about what ``verification'' means. We present these findings alongside respondents' perceptions of verification criteria and account characteristics that are \textit{required} of verified accounts in general, and across three specific social media platforms (\textcolor{blue}{RQ1}). We then perform statistical analyses of any discrepancies in these perceptions across platforms, and identify statistically significant differences between respondents' perceptions of platform-guaranteed verification assurances and generally acceptable verification criteria (\textcolor{blue}{RQ2}). Next, we specifically evaluate Twitter's taxonomy of verification marks, including its gray and gold checks in addition to Twitter Blue check marks (\textcolor{blue}{RQ3}). We conclude with an analysis of possible respondent age and digital literacy interactions with perceptions of verification marks. ({\textcolor{blue}{RQ4}). 

\begin{figure*}[h!]
\centering
\includegraphics[scale=0.61]{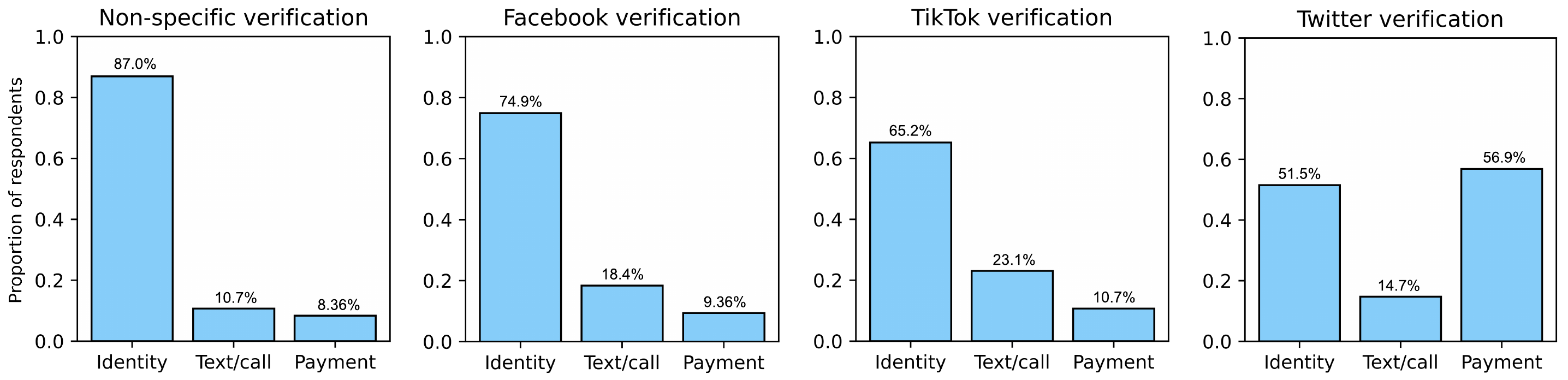}
\caption{Perceived criteria for verification (Full study). Survey respondents' perceptions of criteria for verification (identity verification, confirmation of receipt of text or call, or payment) on Facebook, Twitter, TikTok, and in general.}
\label{fig:verif_reqs}
\end{figure*}

\subsubsection{Understanding Existing Verification Marks}
\textbf{Defining ``verification'' and identifying verification criteria.}
Before addressing perceptions of \textit{verification criteria}, we gathered free responses to an open-ended question about respondent perceptions of \textit{verification}; a free response question format allowed us to avoid prompting effects and the limitations of a structured question. To that end, we asked respondents to write responses to the following question: ``In general, what does a `verified' badge on an account mean to you?'' We performed inductive coding on respondents' answers to this question. Our methods and full inductive coding taxonomy are available in~\Cref{sec:verif_taxon}. We highlight some interesting observations from our coding: 

\textbf{About 61\% (181/299) of respondents} stated that ``verification'' meant that an account-holder was whom they claimed to be---these responses frequently made mention of spam accounts, bots, and celebrities. The purpose of verification is twofold, according to responses in this category: 1) verification of account-holder identity, and 2) prevention of bots and imitators. A representative response is as follows: ``[verified users] are famous and can have people that pretend to be them.'' A few terms synonymous with ``verified'' made frequent appearances, including ``authentic,'' ``legitimate,'' and ``real.'' One respondent, acknowledging the notability of verified accounts, wrote: ``Verification in my opinion should be given to very prominent people and not just anyone. So the person behind the account should be taken into consideration.'' Existing research supports this finding---in a study from 2019, Vaidya et al. found that the majority of respondents to a \textit{structured}-response question identified ``verification'' as a confirmation of account-holder identity~\cite{vaidya-2019-verified-credibility}. 

\begin{table}[t]
\small
\label{table:verif_meaning_taxon}
\begin{tabular}{lr}
\toprule
Response & Proportion \\
\midrule
        Account holder identity has been verified & 0.605 \\ 
        Account belongs to a notable entity & 0.157 \\ 
        Account was vetted by platform & 0.151 \\ 
        Account was paid for & 0.0870 \\ 
        Account is credible/factually reliable & 0.0736\\ 
        Account is influential/has high followership & 0.0401 \\
        No meaning & 0.0800\\
        Uncertain & 0.0172 \\
        N/A (no response, or nonsense response) & 0.0133 \\
\bottomrule
\caption{Respondent perceptions of social media ``verification,'' by proportion of survey responses. Taxonomy produced via inductive coding of free response answers. Note that a single response might fall into multiple categories.}
\end{tabular}
\end{table}

\textbf{About 15\% (45/299) of respondents} stated that verified status meant that a social media platform had vetted an account, and had applied its own set of criteria in order to do so. These responses did not directly address the \textit{meaning} of verification, focusing instead on the verification \textit{process}. They distinguished between actual guarantees of identity verification and platform-specific promises. One respondent wrote: ``[The user] went through whatever process to become verified on a particular platform.  It doesn’t mean as much as it used to.  Especially on Twitter after Musk took it over.'' 

\textbf{About 9\% (26/299) of respondents} explicitly mentioned payment in their response---and some respondents stated that verification meant very little or nothing at all to them because of Twitter's paid subscription model. Per one survey respondent: ``If it’s a legacy verified badge, it means that the person is legit and not a troll or nobody. If it’s a paid for badge, it means the person is a sucker.'' 

\textbf{About 7\% (22/299) of respondents} mentioned trustworthiness, credibility, or factuality in their response. This type of response, while infrequent, is concerning because identity verification and source credibility are distinct concepts. As an example, one respondent wrote that verification confirms ``the info on the account is from a reputable organization that only deal in facts.'' The events of the past few years have demonstrated that verified accounts in categories that people commonly associate with verification, such as celebrities and politicians, are not always reliable sources of information. This result is consistent with the findings in Vaidya et al.~\cite{vaidya-2019-verified-credibility}. 
        
The largest response category---the first group listed above---understands \textit{verification} to be \textit{identity verification}: confirmed association of an online persona with an offline person. A full taxonomy is available in~\Cref{table:verif_meaning_taxon}.

With this context, we now address the remainder of \textcolor{blue}{RQ1}, regarding user perceptions of existing verification criteria across different social media platforms. We analyze responses to a structured survey question: Respondents were presented with 9 different criteria for verification and asked to select the criteria which they believe are required for verification across different platforms and in a non-specific case. We highlight respondent perceptions of three of these criteria---identity verification, text/call confirmation, and payment---in~\Cref{fig:verif_reqs}. We make a few observations: 

\textbf{The vast majority of respondents (87\%, or 260/299)} indicate that confirmation of identity is, in general, a requirement to obtain verified status on social media platforms. This result is a greater share of respondents than gave a similar answer to the free response question, suggesting some lack of familiarity with and uncertainty about verification requirements.

\textbf{Fifty-two percent (154/299) of respondents}  believe that confirmation of identity is required for a Twitter account to receive a blue check. We posed the question about current requirements for obtaining a blue check, so these responses are factually mistaken about Twitter's practices. 

\textbf{\textit{More} respondents (57\%, or 170/299)} believe that payment is required for a Twitter account to receive a blue check (i.e., the account must be subscribed to Twitter Blue). This result shows that respondents are generally aware of Twitter's transition to paid verification, while remaining unaware that verification no longer involves identity confirmation.

\textbf{There is a discrepancy} between perceived \textit{non-specific} verification requirements and perceived \textit{platform-specific} requirements. This result suggests possible uncertainty about platform requirements and lack of trust in platforms. We discuss this gap in greater detail in the next section.

\begin{table}[t]
\small
    \centering
        \centering
        \begin{tabular}{lrr}
        \toprule
            Platform & Identity & Payment \\
            \midrule
             Non-specific & 260 (0.870) & 25 (0.0836) \\
            Facebook & 224 (0.749) & 28 (0.0936) \\
            TikTok & 195 (0.652) & 32 (0.107) \\ 
            Twitter & 154 (0.515) & 170 (0.569) \\
        \bottomrule
        \end{tabular}
    \caption{Perceived requirements for verification: identity confirmation (``Identity'') and paid enrollment (``Payment''). We report raw counts and proportions (\textit{n}~=~299).}
    \label{table:sig_verif_full}
\end{table}

\subsubsection{Verification Assurances: Expectations vs. Reality}    
Toward understanding the gap between respondent expectations of verification assurances and verification assurances in practice (\textcolor{blue}{RQ2}), we focus on answer choices \textbf{``Identity''} and \textbf{``Payment''} in~\Cref{fig:verif_reqs}, corresponding to ``the platform confirms identity of person or organization operating the account'' and ``the account has a paid subscription,'' respectively. We consider respondent perceptions of these criteria for specific platforms and in general (labeled ``Non-specific'' in~\Cref{table:sig_verif_full}). The null hypothesis for our binomial test analysis states that there is no statistically significant difference between the number of respondents who perceive identity verification or payment to be criteria for verification \textit{in a platform-specific case} versus \textit{in general}. 

 We find that significantly fewer respondents perceive \textit{identity} verification to be a requirement for Facebook, TikTok, and Twitter verification, compared to our general case (\textit{p}~<~0.05 for all three platforms, 95\% CI); this contrast is most visible in our Twitter data. For paid enrollment, there is a statistically insignificant difference between Facebook and TikTok and the general case (\textit{p}-value for Facebook is 0.306, and 0.0916 for TikTok). For Twitter, however, \textit{p}~<~0.05, indicating that significantly more respondents view paid enrollment as a requirement for verification. Across all three platforms (and for Twitter, in particular), respondents appear to perceive verification to be less rigorous than what they understand verification, in general, to be.

\subsubsection{Evaluating Twitter Blue's Verification Process}
\label{sec:twitter_blue}
Here, we discuss user perceptions of Twitter's verification processes (\textcolor{blue}{RQ3}). Again, we purposefully present a review of free response questions adjacent to our analysis of a series of multiple-choice questions in order to surface trends not baked into a structured problem statement. 

In detail: one of our survey questions (Q32) described the Twitter Blue program without naming it explicitly, then asked respondents if the criteria described in the problem statement constituted adequate verification (Q35). A majority claimed that the verification criteria, as described, appeared insufficient: of 299 total survey respondents, 29\% (86) felt that the described criteria were ``somewhat insufficient'' and 27\% (82) felt that the criteria were ``very insufficient.'' 11\% (32) submitted neutral responses---``neither insufficient nor sufficient.'' Finally, 26\% (78) felt that the verification requirements were ``sufficient,'' and 7\% (21) felt that they were ``very sufficient.''

If a respondent felt that the described verification process was somewhat or very insufficient, we asked them to describe, in a free response, what part(s) of the process appeared lacking. We received 168 responses in total and inductively coded them as follows: a researcher read through all responses, building the codebook as she noticed recurring response types; ultimately, several response types emerged, though many responses belonged to more than one category. Most responses expressed at least one of the following four opinions: 

\textbf{About 65\% (110/168) of respondents} said that the verification criteria described in the problem statement do not actually confirm the account-holder's \textit{identity}. Per one respondent: ``[The user's] identity isn't confirmed and you can just pay to be verified.'' (Note that many responses in this category also belonged to the next one in our taxonomy.) As discussed previously, about 61\% of respondents mentioned identity verification as a criterion for verification in an \textit{unstructured}, free-response setting. The 4\% increase in the proportion of responses of this kind was possibly due to the explicit mention of identity verification in the \textit{structured} multiple choice problem statement. 

\textbf{About 12\% (20/168) of respondents} said that verification should \textit{not} be subscription-based or otherwise for sale. A response to this effect:  ``Any service that wants you to pay isn't sufficient [verification].''

\textbf{About 5\% (9/168) of respondents} said that confirmation of receipt of a phone call or text at a user-provided number does not constitute actual verification of account holder identity. Respondents pointed out that such a verification measure would be fairly easy to dodge: ``calling a number [users] have provided, it could be anyone.''

\textbf{About 4\% (7/168) of respondents} said that verification marks should be reserved for notable accounts, such as those belonging to celebrities, high-profile organizations, and news outlets. According to one respondent,  verification ``should only be reserved for public figures, businesses, organizations and other important entities, not for everyone in general.'' More than half of survey respondents ranked news outlets, journalists, and business organizations among the entities they most commonly follow on social media (see~\Cref{app:accts_followed}).

These free response results provide a summary of perceived shortcomings of Twitter Blue's verification standards. We elaborate upon possible solutions to the problem posed in the third category of responses, regarding phone/text confirmation as a proxy for identity verification, in~\Cref{sec:improvements}.

\textbf{Evaluating Twitter's other checks.} In addition to Twitter Blue's blue check marks, accounts might also receive a gold or gray check mark (\Cref{table:verif_marks}). As stated in Twitter's documentation, gold checks denote businesses and other organizations, while  gray checks denote a government component or official. In practice, however, these guidelines have not been evenly applied (see~\Cref{sec:verif-marks}). Our survey respondents expressed a good deal of confusion about the meaning of these checks. Of 299 survey respondents, 80\% (239) stated that they do not know what a gold check mark means, and 87\% (259) stated that they do not know what a gray check means. Nine percent (26) correctly stated that gold checks identify businesses, and 5\% (15) correctly stated that gray checks identify governmental organizations and political figures. 

Notably, the remaining respondents (about 24 people, for both gold and grey checks) mostly fell into two categories: those who believed that gray or gold check marks identified accounts verified by Twitter under its previous non-paid model, and those who believed that gold and gray check marks were different paid subscription tiers.\footnote{After the conclusion of our survey, Twitter launched a paid version of gold checks for Verified Organizations~\cite{tw-gold-verif-2023}.} These intermediary responses, which show some degree of awareness of a change in the taxonomy of Twitter verification marks, were a small but significant portion of survey responses. They also lend themselves to two opposite conclusions, each erroneous in its own way: 1) that gold and gray check marks are legacy-verified accounts, and should be trusted; or 2) that gold and gray check marks are paid subscription tiers, perhaps above Twitter Blue, and should not be trusted for that reason. The first line of reasoning is erroneous because gold and gray checks do \textit{not} denote legacy-verified checks---they denote verified businesses and government organizations and officials, respectively. The second line of reasoning is erroneous because gold and gray checks were not subscription-based, at the time of survey deployment. \\

\begin{table}[t]
\small
\centering
\begin{tabular}{ l c c } 
\toprule
Variable  & ID (Twitter) & ID (General) \\
\midrule
    & 0.993 & 1.05 \\
 Digital Literacy   & (0.958, 1.03) & (0.994, 1.10) \\
    & p~=~0.714 & p~=~0.0855 \\[0.25cm]
    & 1.00 & 1.01 \\ 
 Age   & (0.988, 1.02) & (0.988, 1.03) \\
    & p~=~0.691 & p~=~0.369 \\[0.25cm]
  & 0.997 & 1.01 \\ 
Usage    & (0.919, 1.08) & (0.894, 1.14) \\
    & p~=~0.933 & p~=~0.863 \\ [0.25cm]
    & 1.04 & 0.776 \\
Aware of Musk Leadership    & (0.510, 2.11) & (0.249, 2.42) \\
    & p~=~0.921 & p~=~0.663 \\
\bottomrule
\end{tabular}
\caption{Logistic regression analysis of demographic variables and perceptions of verification criteria. We report odds ratios, confidence intervals, and uncorrected \textit{p}-values.}
\label{table:reg_table}
\end{table}

\subsubsection{Demographic Interactions with Perceptions of Verification Marks}
We begin our analysis of demographic interactions (\textcolor{blue}{RQ4}) by conducting a simple bivariate correlation analysis. The variables that we consider are respondent digital literacy, age, frequency of social media usage, perception of identity confirmation as a verification requirement on Twitter, and perception of identity confirmation as a verification requirement in general for social media platforms. \Cref{table:tech_lit_percep} presents the results. The analysis shows a modest (and not statistically significant) negative correlation between digital literacy and identity confirmation as a verification requirement on Twitter (i.e., respondents with greater digital literacy were more likely to accurately respond that Twitter blue checks do not confirm identity). The analysis shows a modest positive (and again not statistically significant) correlation between age and identity confirmation as a requirement (i.e., respondents who were older were more likely to misunderstand Twitter Blue). The results include a statistically significant negative correlation between digital literacy and age, and a significant positive correlation between digital literacy and social media usage.

Next, we conduct a multivariate logistic regression to understand joint interactions among demographics. Our independent variables are respondent self-reported digital literacy, age, Twitter usage, and awareness of Elon Musk's leadership. Our dependent variables are respondent perceptions of identity verification as a requirement for account verification \textit{on Twitter} and \textit{in general}. We use logistic regression because the dependent variable is a binary outcome (perceptions of identity as \textit{required} or \textit{not required} for verification). Our null hypothesis is that no relationship exists between any of our demographic variables and perceptions of verification criteria both on Twitter and in general (i.e., odds ratios equal to 1). A summary of our findings can be found in~\Cref{table:reg_table}. We again find a modest and not statistically significant relationship between digital literacy and perceptions of verification. The relationship between age and perceptions, however, disappears in the multivariate analysis. One possibility is that no true relationship exists. Another is that digital literacy ``explains away'' age as a correlate of account verification perceptions, because age and digital literacy are correlated.

\begin{table}[t]
\small
    \centering
        \begin{tabular}{ l r r r r}
        \toprule
            & 1\phantom{${}^{\ast}$} & 2 & 3 & 4 \\
            \midrule
            1. Digital literacy &  &  &  &   \\ 
            2. Age  & -0.232$^\ast$ &  &  &   \\ 
            3. Usage & 0.103$^\ast$ & -0.229 &  &   \\
            4. ID (Twitter)  & -0.028\phantom{${}^{\ast}$} & 0.030 & -0.012 &   \\
            5. ID (General) & 0.090\phantom{${}^{\ast}$} & 0.032 & 0.004 & 0.198  \\
        \bottomrule
        \end{tabular}
    \caption{We calculate Pearson's \textit{r} for all pairwise combinations of the following variables: 1) self-reported digital literacy, 2) respondent age, 3) frequency of respondent social media usage, 4) respondent perceptions of identity verification as a requirement for verification on \textit{Twitter} (denoted ``ID (Twitter)''), and 5) respondent perceptions of identity verification as a requirement for account verification on social media platforms \textit{in general} (denoted ``ID (General)'').} 
    \vspace{4pt}
    $^\ast$ indicates significance at the $p < 0.05$ level.
    \label{table:tech_lit_percep}
\end{table}

\begin{table}[t]
\small
    \centering
    \begin{tabular}{rrrr}
    \toprule
        Digi. lit. score & \# Respondents & In general & On Twitter \\
        \midrule
        17-35 & 77 & 0.792 & 0.481\\
        36-41 & 75 & 0.947 & 0.653 \\
        42-45 & 81 & 0.901 & 0.457\\
        46-50 & 62 & 0.855 & 0.468\\
    \bottomrule
    \end{tabular}
    \caption{Proportion of respondents who believe that identity verification is a requirement for account verification \textit{in general} and \textit{on Twitter}. ``Digi. lit. score'' corresponds to self-reported digital literacy scores, by quartile.}
    \label{table:techlit_percep}
\end{table}

\paragraph{Digital literacy and age.}
In order to calculate a composite per-respondent digital literacy score, we asked respondents to rate their own understanding of ten different terms pertaining to technology and the Internet on a 5-point Likert scale. The prompt included a meaningless dummy term, ``Proxypod,'' to capture possible overconfidence in understanding. The terms that we used were a subsample of those that appeared in a classic digital literacy survey instrument designed by Hargittai and later updated by Munger and Guess~\cite{hargittai-survey-2005, guess-munger-diglit-2022}. We assigned a numerical value to each self-rating (with ``no understanding''~=~1 and ``full understanding''~=~5) and summed respondents' self-ratings over all ten queried terms, with 50 being the highest possible score and 10 being the lowest. 

Of all the independent variables in our regression analysis, digital literacy has the strongest beneficial relationship for understanding Twitter account verification. We observe slightly \textit{decreased} odds of perceiving identity verification as a criterion for verification \textit{on Twitter} among more tech literate respondents (OR = 0.993, 95\% CI [0.958, 1.03]). Interestingly, more tech literate respondents also had higher odds of perceiving identify verification to be required for verification \textit{in a general case} (OR = 1.05, 95\% CI [0.994, 1.10)]). These results suggest that greater digital literacy improves a social media user's odds of knowing what verification assurances platforms should---and do---guarantee. While we do not observe significant age-induced effects on perceptions of Twitter verification in our logistic regression analysis, there is a weakly negative correlation between age and digital literacy (reported in~\Cref{table:tech_lit_percep}), suggesting that older respondents are more likely to self-report lower levels of digital literacy. 

On average, older users rated their understanding of technology terms lower than did their younger counterparts. This finding is tempered by the fact that younger respondents were also more likely to self-report some degree of understanding of ``Proxypod'' (i.e., they gave themselves a score of 2 or greater): about 25\% (13/51) of respondents in the 18-27 age group claimed some understanding of the dummy term, compared to 13\% (7/53), 19\% (9/48), 22\% (11/49), and 12\% (11/91) for the 28-37, 38-47, 48-57, and 58-100 age groups, respectively. We offer further commentary on the potential pitfalls of self-reported digital literacy scores in~\Cref{sec:limitations-survey}.

\paragraph{Awareness of Musk's leadership and Twitter usage.} 
We asked survey takers: ``Who is the current leader of Twitter?''\footnote{ We opted for this wording because Musk's official relationship to Twitter is in flux. In April 2023, he stated that his ``dog is the CEO of Twitter''~\cite{elon_dog_ceo}. In May 2023, he announced Linda Yaccarino as the new CEO of Twitter \cite{Grantham_2023}.} Correct answers were recorded as 1s and incorrect answers were recorded as 0s. Per our regression analysis, there may be a modest and counterintuitive relationship between awareness of Musk's leadership and perceptions of verification marks (i.e., familiarity with the leadership increasing misperceptions about verification). This result is not statistically significant, however.

We also asked respondents to report their Twitter usage on a weekly basis, with the lowest frequency being ``never'' and the highest frequency being ``every day.'' We observe a slight correlation between greater use of Twitter and accurate understanding of account verification, though again, the result is not statistically significant.

\subsection{Limitations}
\label{sec:limitations-survey}

\textbf{Recruiting biases.}
We recruited survey respondents through Prolific, rather than traditional means of assembling a panel (e.g., telephone calls or postcards). This recruiting method may have introduced relevant biases that are not accounted for by the representative demographic factors that we applied. In particular, it is possible that the relationship between age and verification perceptions observed among Prolific respondents is a lower bound for the general population, as Prolific respondents are likely more tech-savvy. These respondents are at minimum capable of locating, navigating, and completing a paid online survey, and often participants on Prolific (and similar platforms) are frequent survey takers. Recent research has shown that online survey platforms might select for respondents with better-than-average digital skills~\cite{guess-munger-diglit-2022}. Follow-on work could address this limitation by using alternative recruiting methods.

\textbf{Respondent attentiveness and news awareness.}
We included one attention check question and two news awareness questions in our survey. The attention check question was intended to identify respondents who were not carefully reading prompts, and the news awareness questions both served the same goal and provided additional data for analysis.

As noted earlier, we excluded one participant who gave a plainly incorrect answer to a news awareness question about the current U.S. President, leaving 299 respondents in our study. We used the other news awareness question, about Twitter's leadership, in the earlier bivariate correlation analysis (\Cref{table:reg_table}). While the correct answer was ``Elon Musk,'' we credited any plausible answer as sufficient to pass the question.

Unfortunately, we found that many of the remaining participants (88\%, 263/299), failed the attention check question despite having otherwise valid survey responses. In a manual review of answers, especially free response prompts, these participants clearly demonstrated that they were paying attention to the study. We interpreted this result as an indication that our attention check question, which instructed participants to disregard a prompt and select a particular response, was unintuitive and confusing. We provide a robustness analysis in~\Cref{sec:attn_check} showing that the attention check question had little relation to perceptions of Twitter's verification changes.

\textbf{``Required'' vs. ``typical'' characteristics of verified accounts.} In our survey, we intended to separately gauge two aspects of verification. First, what do people believe platforms \textit{mandate} to obtain verification? This survey component allows us to understand if respondent perceptions of verification marks align with platform practices and the threat model. We use this component of the survey in the analysis above. Second, we sought to understand: what attributes are perceived to be \textit{common} among verified accounts? We intended to directly compare these results to our measurements, to understand whether people's perceptions about the types of accounts that gain verification match the actual types of accounts obtaining verification. Unfortunately, we found that the prompt distinction was too subtle and participants generally gave identical answers. As such, we omit responses to our ``typical'' prompts from the analysis that we present here. 

\textbf{Self-reported digital literacy scores.} 
Two common approaches to measuring digital literacy are 1) requesting respondents to self-report their familiarity with terminology, and 2) observing and measuring respondent performance on a task requiring tech skills \cite{hargittai-skills-2002}. We adopt the first approach in our study. Based on responses to our digital literacy dummy term, we observe some likely inflation of self-reported understanding of technology-related terminology, particularly among younger users. This method for measuring digital literacy necessarily requires truthful responses from survey takers, and some bias is to be expected in respondents' subjective self-evaluations of their own digital literacy. A possible explanation for these inflated scores is that ``Proxypod,'' our choice of dummy term, might have sounded  similar to legitimate technical terminology (e.g., proxy servers). 

\section{Measurement Study: Longitudinal Analysis of Verified Accounts}
\label{sec:measurement}
Our survey study investigated user perception of account verification on social media platforms. This complementary study aimed to understand the reality of which accounts currently have such verification indicators via a measurement study of Twitter accounts with varying verification status. In particular, we are interested in analyzing the sets of users that possessed a blue check mark on Twitter before the platform's changes and those marked with the same verification indicator afterward (with different semantics). We conduct a data-driven analysis of tweet datasets in order to answer the questions posed in~\Cref{sec:data-driven-analysis-rq}.

\textcolor{blue}{RQ5.} We analyzed account attributes for Blue subscribers and legacy verified accounts, which use the same blue check mark indicator. We also manually coded account types for a random sample of 157 Blue accounts, 173 legacy verified accounts, as well as 149 randomly sampled accounts to use as a control.

\textcolor{blue}{RQ6.} In order to identify trends within the content being published by different sets of verified accounts, for 157 Blue accounts and 173 legacy verified accounts, we inductively coded recent tweet content to surface differences between the two groups of accounts. We apply these same codes to a sample of 149 randomly sampled to serve as a control.

\begin{table}[t]
\small
\centering
\begin{tabular}{ clr } 
\toprule
Legacy verified & New verified type & Count \\
\midrule
False & Blue & 23.3K \\ 
 & Business & 288 \\ 
 & Government & 14 \\
 & None & 2.81M \\
\midrule
True & Blue & 438 \\ 
 & Business & 1.93K \\ 
 & Government & 353 \\
 & None & 19.8K \\
\bottomrule
\end{tabular}
\caption{Number of user accounts in each Twitter verified status group. ``Business'' and ``Government'' verified accounts bear gold and grey check marks, respectively.}
\label{table:verified_user_count}
\end{table}
\subsection{Dataset Curation}
On November 8, 2022, we randomly sampled over 15 million tweets published in the month of October 2022 using the Twitter API for Academic Research.\footnote{Academic access to the Twitter API has been discontinued as of April 29, 2023~\cite{Floridi_Morley_Juneja}.} We collected random tweets by sampling 33.6k timestamps uniformly at random from the month of October, and pulling the 500 most recent English language tweets at those timestamps using the full-archive search endpoint. We extracted account data for all accounts represented by tweets in this sample. This data contained all fields provided by the Twitter API, including username, geographical location, verified status (from the \texttt{verified} field), and Twitter ID.

On January 10, 2023, we checked the verified status of the accounts for which we extracted account data. In total, after filtering out accounts which were deleted, we collected verified status changes for 2.85 million accounts that tweeted in October. A new \texttt{verified\_type} field appeared in the data, corresponding to the type of verification mark applied to the account, if applicable. This update was reported in Twitter's API changelog on December 21, 2022~\cite{twitter_platform_changelog}. Different \texttt{verified\_type} responses included \texttt{business} (for accounts with gold checkmarks indicating a verified business), \texttt{blue} (for accounts with a paid subscription to Twitter Blue), \texttt{government} (for government agencies or officials), or \texttt{none} (used for legacy verified accounts and non-verified accounts). Using the \texttt{verified} field and the new \texttt{verified\_type} field, we could deduce a range of actions taken by the user: e.g., an account labeled as \texttt{verified = false} in our initial October sample and, later, as \texttt{verified\_type = blue} in our January verification status check is an account that was unverified and later subscribed to Twitter Blue.

There were a handful of accounts whose legacy \texttt{verified} status changed in this time period. Since the reason for legacy verification churn was unclear during this period of time, and since this case made up a negligible portion of the dataset (less than .003\%), we only considered accounts whose legacy \texttt{verified} status remained the same between October and January (count of accounts per bucket is described in~\Cref{table:verified_user_count}). 

At the time we conducted this measurement study, Blue verified accounts used the same verification indicator that legacy verified accounts  displayed, only distinguishable by first visiting the account profile and then clicking the verification indicator.\footnote{This user interface has repeatedly changed since, and at the time of publication does not distinguish between Blue and legacy verification.} Our goal is to analyze practical differences between groups of legacy verified accounts and Blue verified accounts that use this indicator. Thus, we consider three groups of accounts: \textbf{Blue verified accounts} that were not legacy verified, \textbf{legacy verified accounts} that were not Blue verified, and \textbf{control accounts} which are sampled uniformly at random from all accounts (control sampling was agnostic to verification status).

We sampled 200 accounts uniformly at random from each of these groups. We then filtered out any accounts that were no longer accessible (i.e., the account was either protected or deleted) and accounts that did not tweet mostly in English. We pulled the remaining accounts' most recent tweets, up to the maximum 3,200 allowed by the Twitter API. 

We manually coded these user accounts, referencing their recent tweets, timeline, and description in order to inductively produce a taxonomy of account types and common themes in tweet content for each account.

\subsection{Account and Content Analysis}
Our approach to qualitative tweet analysis focused on structural, conceptual coding via inductive analysis \cite{chandra_coding_2019,saldana_coding_2021}. Concretely, two researchers first studied and discussed a smaller sample of accounts from our datasets, and grouped together high-level themes that emerged from the data. The researchers then used these themes as a codebook to code another small sample of accounts, and iterated between discussion and coding more samples until the codebook stabilized. For the account taxonomy, we started out with the sections presented to survey participants (as in~\Cref{app:accts_followed}), and added other common account types that emerged from inductive coding.

Using these methods, we developed codes for account types and tweet content that were salient to our research questions. First, we marked accounts that were not tweeting primarily in English, as English was the only common native language of the coding researchers. Though we sampled these accounts from English language tweets, Twitter's language metadata is not always accurate, and many accounts tweet in multiple languages. Then, we classified the account type and content according to the codebook in~\Cref{sec:tweet-codebook}.

For content analysis, in light of the varied content published by Twitter accounts, we focused on developing codes for answering \textcolor{blue}{RQ6} by surfacing content trends that varied substantially between legacy and Blue verified accounts. For instance, we identified a stark difference between the number of legacy and Blue verified accounts that emphasized marketing for a cryptocurrency or NFT project. We also identified differences in political leanings between the two account groups and found that, in our Twitter Blue dataset, a handful of accounts heavily retweeted Elon Musk. We identified these as codes that were salient to the content analysis, and we also used these codes to perform a control analysis on non-verified accounts. The full content codebook is described in~\Cref{sec:tweet-codebook}.

Finally, in order to measure inter-rater reliability, each researcher redundantly coded 20\% of the total samples and did not indicate which were duplicated in the other's sample. Using these 20\% overlapping samples, we calculate Cohen's kappa to measure inter-rater reliability between the two coders. This statistic was 91.5\% for account categorization and 82.3\% for content tagging, which indicates a reasonable amount of agreement between our coders.

\begin{figure}
\includegraphics[width=0.99\textwidth]{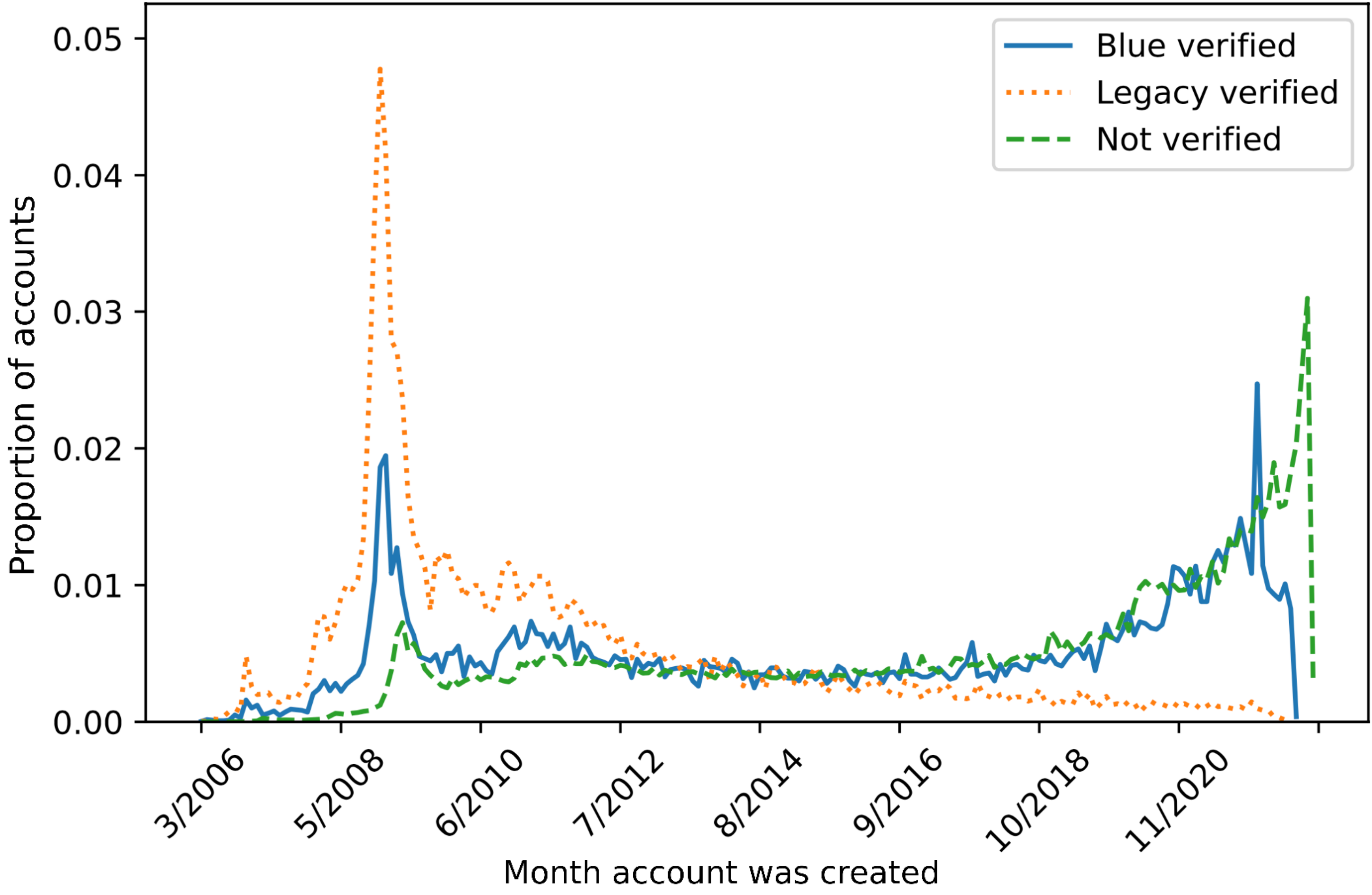}
\caption{Distribution of accounts registered at a particular date, binned by months. The distribution of Blue verified accounts skews more recent than does the distribution of legacy verified accounts, and more closely resembles the distribution of non-verified account registration dates.}
\label{fig:accts_registered}
\end{figure}

\subsection{Limitations}

\textbf{Inconsistent geographic or linguistic dataset boundaries.} There may be regional, linguistic, or cultural biases in this analysis. Since the researchers conducting the analysis only shared English as a common native language, we only sampled English-language tweets as determined by Twitter metadata, which is also imperfect. In addition, when we queried verification status again in January 2023, Twitter Blue was only available in certain geographical regions: Australia, Canada, New Zealand, the U.K., and the U.S. However, legacy verification was available for many more geographical regions in which accounts tweet in English. We did not perform any location filtering in our analysis.

\textbf{Bias towards user accounts that tweet more frequently.} Our random sampling method began with a random sample of tweets. As a consequence, the method was biased towards accounts that tweeted more frequently in October 2022.

\textbf{Inherent limitations in political coding, focusing on the U.S. two-party system.} We only categorize political leaning if an account is consistently in support of or opposition to a particular party's candidates within the U.S. two-party system. We recognize this excludes a large portion of political discourse, since we only code the leanings of highly partisan accounts using a clear but narrow political taxonomy. For instance, we did not code event- or issue-based stances, and we also did not code for political contexts outside of the U.S. Our analysis aims to capture larger trends and leanings, and does not purport to holistically capture the nature of online political discourse. We acknowledge that this method may also introduce bias to our analysis.

\subsection{Ethical Considerations}

There are many ethical implications for the analysis of public social media data, including Twitter data. We sourced our data from the Twitter API for Academic Research, which only returns public data. Prior work has demonstrated that some Twitter users are unaware of and not accepting of use of their public content in academic research~\cite{twitter_research_ethics}. To mitigate any possible risks to users whose data we analyze, we sampled from a large pool of accounts and are only publishing our methods and aggregate results. We do not believe it would be possible to re-identify any user from the published components of this research. Furthermore, in this study, we analyze Twitter posts to assign categories that some users may find sensitive (e.g., political views). We took the further steps of removing account IDs, usernames, and tweet IDs from our coded dataset and stored the Twitter data that we analyzed on encrypted and access controlled devices.

\subsection{Verified Account Analysis Results}

\begin{figure}[t]
\includegraphics[width=0.94\textwidth]{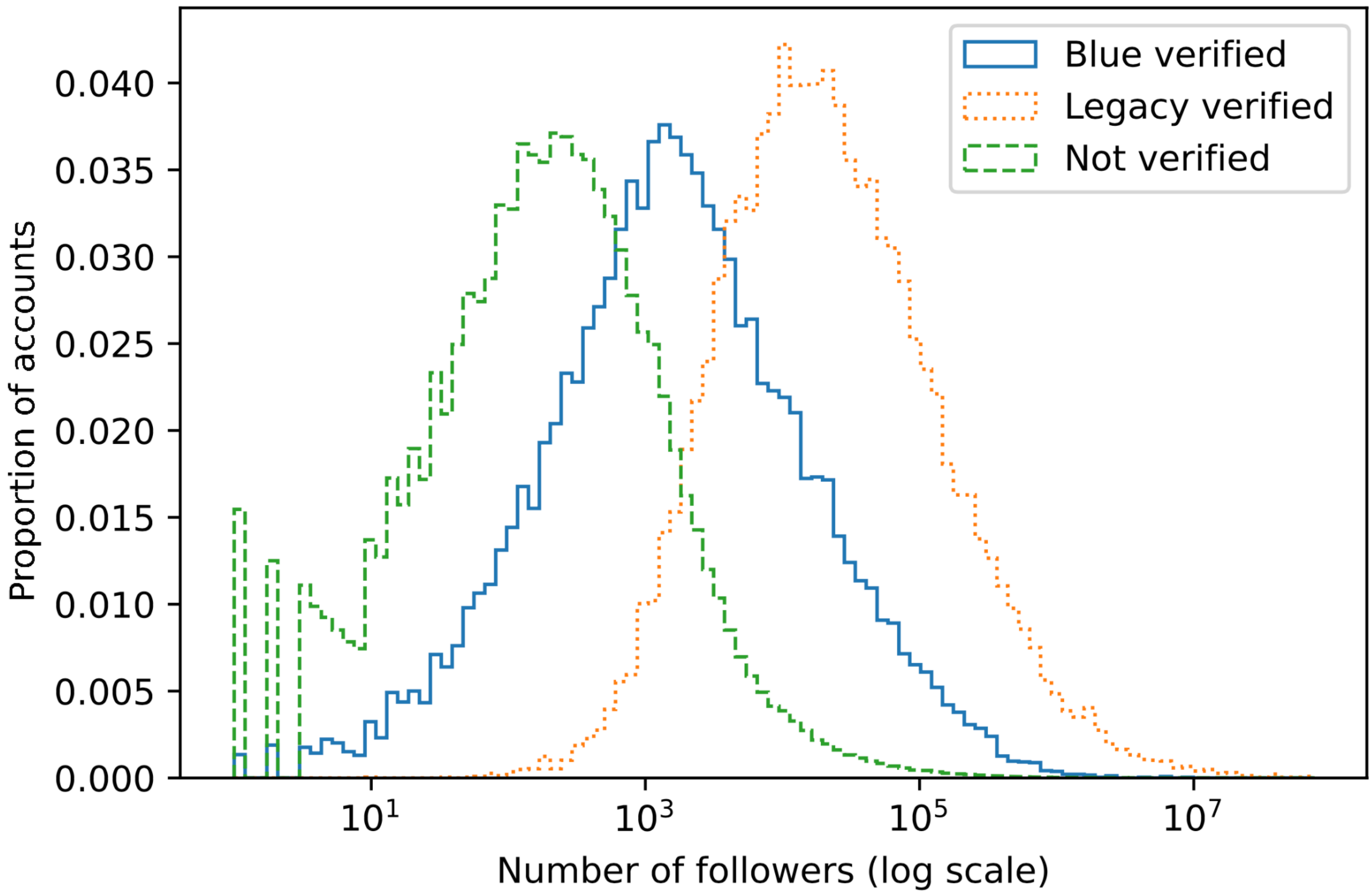}
\caption{Distribution of follower counts (log scale) for accounts with various verification status. That legacy verified accounts have the most followers is somewhat expected, as significant platform presence (top .05\% follower count in their geographic region) was a criterion for many legacy verification categories.}
\label{fig:follower_count}
\end{figure}

\textbf{Account age and follower count.}
The distribution of legacy verified accounts skews much older than the distributions of Blue verified or non-verified accounts; in general, legacy verified accounts also have more followers. This difference in follower count is expected, since one criterion for many legacy verification categories was to be in the top 0.05\% followed or mentioned 
accounts in the user's local region. Account ages for Blue accounts are similar to those of non-verified accounts, as can be seen in~\Cref{fig:accts_registered}. Follower count distributions are displayed in~\Cref{fig:follower_count}.

\textbf{Account composition.}
Twitter Blue account composition more closely resembles the account composition of the control. The disappearance of government Twitter accounts from the Blue sample is also to be expected, since government accounts now use the Twitter gray verification indicator. Journalist and organizational accounts in the legacy verified group are no longer present in Twitter Blue, which is largely made up of personal accounts. Many Twitter Blue accounts were also promoting small businesses (tagged as businesses), or were owned by smaller content creators (tagged as influencers). The full set of account compositions are displayed in~\Cref{table:account-composition}.

\begin{figure}[t]
\includegraphics[width=\textwidth]{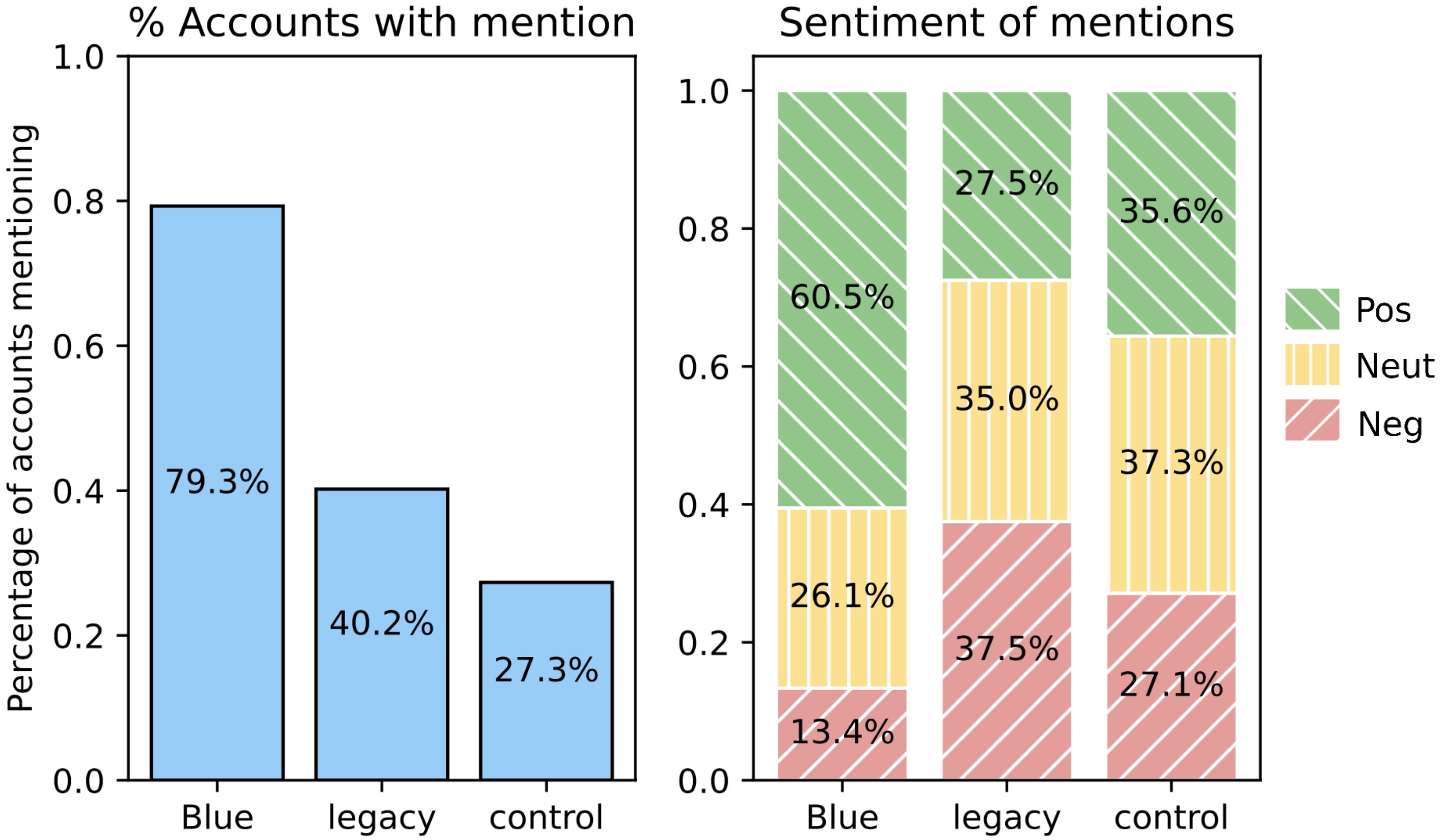}
\caption{\texttt{@elonmusk} mentions and sentiment coding. For each dataset, we pulled any retweets or mentions of \texttt{@elonmusk} from the most recent 3,200 tweets of each account. We then manually coded the sentiment of these tweets. $n=198, 199, 216$ for Blue, legacy, and control, respectively. We describe our taxonomy for sentiment coding in~\Cref{sec:tweet-codebook}.}
\label{fig:elon_sentiment}
\end{figure}

\begin{table}[t]
\small
\begin{tabular}{lrrr}
\toprule
Account type & Blue & Legacy & Control \\
\midrule
Personal account & 0.580 & 0.58 & 0.779 \\ 
Influencer   & 0.191 & 0.254 & 0.0671 \\
Business     & 0.172 & 0.191 & 0.0470 \\
Fan account  & 0.0382  & 0.00  & 0.101 \\ 
Organization & 0.0127  & 0.0809  & 0.0067  \\ 
Journalist   & 0.0064  & 0.364 & 0.00  \\
Government   & 0.00  & 0.0751  & 0.00 \\
Politician   & 0.00  & 0.0289  & 0.00  \\ 
\bottomrule
\end{tabular}
\caption{Account composition for Blue, legacy, and non-verified users. A full description and taxonomy of these account codes can be found in~\Cref{sec:tweet-codebook}. }
\label{table:account-composition}
\end{table}

\begin{figure}[h!]
\includegraphics[width=0.93\textwidth]{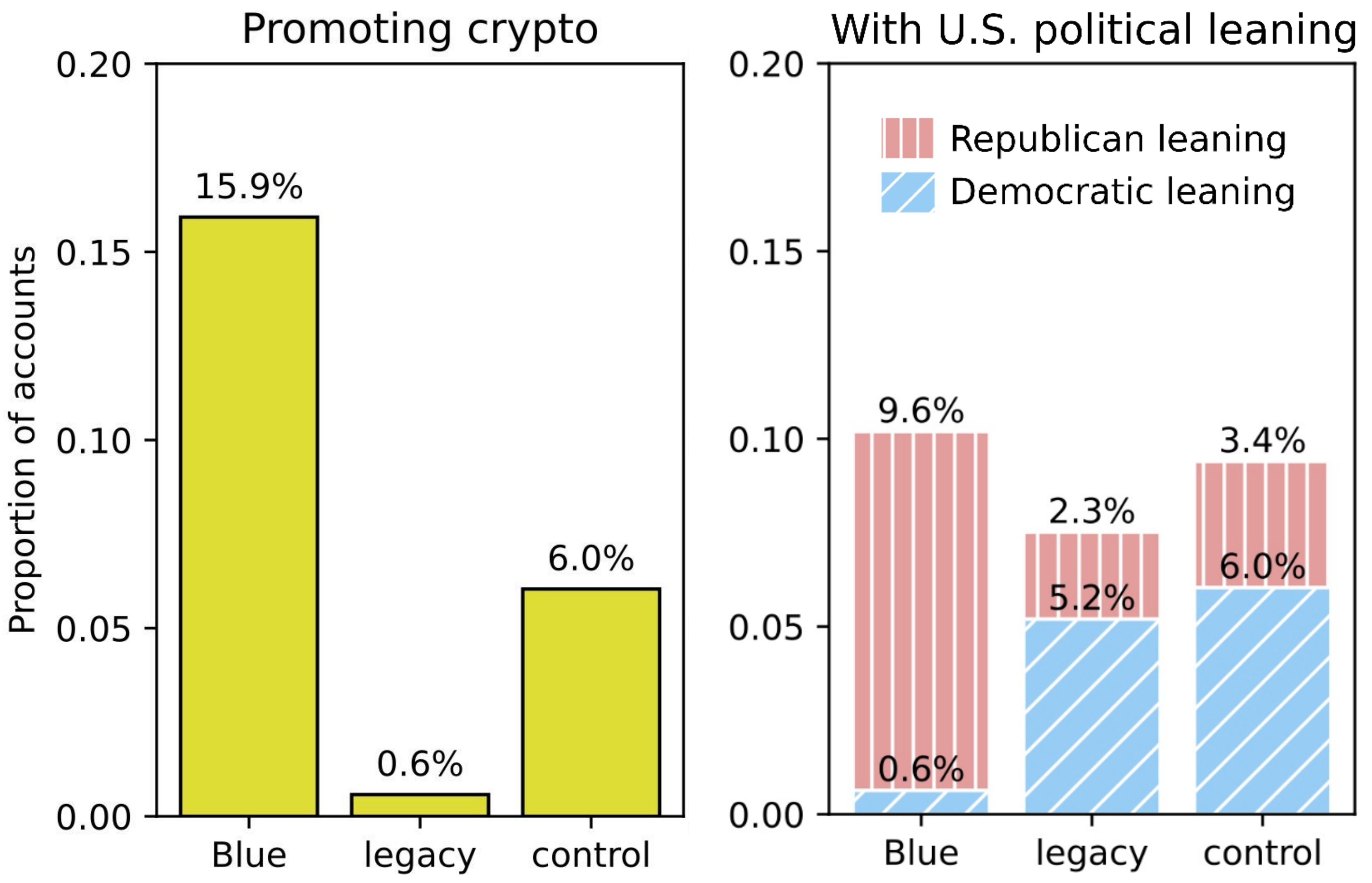}
\caption{Accounts tagged with cryptocurrency or U.S. political leaning content. $n=157,173,149$ for Blue, legacy, and control users, respectively. Using a binomial test, we determined that the differing rate of cryptocurrency promoting accounts between Blue and the other account groups is significant, and that the differing rate of U.S. political leaning between Blue and legacy accounts is significant (with $p < 0.05$). However, there is not a significant difference between the U.S. political leanings of the legacy and control datasets. }
\label{fig:content-analysis}
\end{figure}
\textbf{The prevalence of cryptocurrency promotion in Blue accounts.} As demonstrated in~\Cref{fig:content-analysis}, there is a statistically significant ($p < 0.05$) increase, even from our control, in the relative proportion of Blue accounts promoting sales of a specific cryptocurrency or NFT commodity. Cryptocurrency-related spam is a well-studied and often-reported phenomenon on many social media platforms, particularly on Twitter~\cite{mirtaheri-2021-crypto-manip}. This suggests that cryptocurrency promoters may purchase verification to make their projects appear more legitimate, exploiting user misperceptions about Twitter Blue verification.

\textbf{Discussion about and sentiment towards Elon Musk.}
Finally, we note a significant increase in the number of \texttt{@elonmusk} mentions in the Blue dataset relative to the control. As shown in~\Cref{fig:elon_sentiment}, there is also a statistically significant difference ($p < 0.05$) in the overall sentiment towards Musk. Twitter Blue accounts are generally more favorable towards Musk than control accounts. Legacy verified accounts do not demonstrate a statistically significant attitude towards Musk relative to the control. Twitter Blue subscribers' favorable sentiment towards Musk and political leanings suggests that the Twitter Blue product is closely tied to the CEO's image.

\textbf{Differences in U.S. political leanings.} As demonstrated in~\Cref{fig:content-analysis}, we observed a statistically significant difference in the U.S. political leaning of Twitter Blue accounts relative to legacy verified or control accounts. In 2022, Musk publicly stated his intention to vote Republican in future U.S. elections. His sensationalization of the Twitter acquisition has appealed in particular to right-wing narratives. Our analysis suggests that this has had a strong impact on the composition of Twitter Blue subscribers. However, we did not observe a statistically significant difference in U.S. political leanings between legacy verified accounts and control accounts, which means this effect is unique to Blue subscribers, and that legacy verified accounts were not necessarily more partial to one U.S. political party than the average account that tweeted in October 2022.

\section{Improving Verification on Social Media }
\label{sec:improvements}

Our results demonstrate that there is a clear gap between perceptions of Twitter blue check account verification and the requirements to enroll in Twitter Blue. We also show, more generally, evidence of confusion and uncertainty about social media platform verification practices. Our measurement study demonstrates significant skews in the types of accounts that enroll in paid Twitter verification, including some accounts that may be exploiting misunderstanding of paid verification, highlighting risks of future paid verification offerings.

For Twitter specifically, our results call into question whether the Blue subscription service is consistent with U.S. consumer protection law. A significant mismatch between business practices and consumer understanding (as shown by our survey study), where the discrepancy is material to consumers (e.g., impersonation leading to stock market movements and enabling scams), could run afoul of Section 5 of the Federal Trade Commission Act and similar state laws. Twitter is also under a May 2022 consent order with the FTC, which requires the firm to ``not misrepresent, in any manner, expressly or by implication, the extent to which Respondent maintains and protects the... integrity of Covered Information''\cite{ftc-consent-2022}. There is a plausible argument that Twitter Blue misrepresents account integrity, and the relevant account information is ``Covered Information'' within the meaning of the consent order because verification typically involves a non-public component (e.g., a driver's license).

Below, we offer recommendations towards improving social media account verification practices.

\textbf{Visual separation of subscription and verification status.} The use of the same visual indicator to indicate both a paid subscription and account verification is confusing to users and unnecessary. \footnote{A cynical perspective on Twitter Blue is that the service deliberately conflates these two attributes to improve monetization. From an information security perspective, these two attributes should remain distinct.} Moreover, the accounts that choose to pay for a subscription will not be the accounts most at risk from impersonation and which would give rise to the greatest risk if impersonated. Other platforms, such as Signal, use an additional non-check mark badge next to user icons to indicate that a user has contributed money to the platform~\cite{signal_donor_badge}. We recommend similarly visually distinguishing subscription badges, which indicate a user has paid for certain platform features, from verification badges. We also recommend providing descriptive explanation, in context, of what verification means. If a platform provides verification as part of a subscription, like Twitter Blue or Meta's subsequent offering, that type of verification should be visually distinct as well since the attributes and behavior of paid verified accounts will be so different from other verified accounts that users have established expectations about.

\textbf{Verification should mean rigorous confirmation of identity.} 
After Twitter Blue's initial launch in November of 2022, the firm relaunched Blue in December, with the promise that the new iteration would provide more stringent verification guarantees. In a tweet from November 25, 2022 announcing the December launch, Musk wrote that ``all verified accounts will be manually authenticated before check activates... [p]ainful, but necessary.'' It is unclear if Twitter actually instituted this change and, at any rate, the program remains easy to exploit: a Washington Post journalist trivially spoofed Senator Ed Markey's account by changing the username on an existing account to \texttt{@SenatorEdMarkey}, then purchasing a temporary phone number through T-Mobile. These actions easily skirted Twitter Blue's account age and linked phone number criteria~\cite{fowler-markey-2022}.}\footnote{Twitter Blue's current safeguards are more akin to anti-bot defenses than identity confirmation.}

Verification as confirmation of identity has been a point of consistency in semantics over time and across platforms. Our survey study demonstrates, consist with prior work, that users expect these semantics. Anything short of identity confirmation will be misleading to users. 

One standard for identity verification is the provision of some form of government-issued ID, such as a driver's license, birth certificate, or passport. Users might take a photo of this ID and upload it to the platform for manual verification. We recommend this kind of verification for notable public figures, public officials, or politicians. We caution against requiring ID verification for all types of verified accounts; this has significant privacy implications for accounts belonging to individuals around the world who might need anonymity to avoid offline retaliation, such as journalists, activists, or sex workers. Some users who seek verification may also lack or be unable to obtain a government-issued ID.

As an alternative, online message boards have long employed social proof solutions, which may be effective for high-profile users.  On Reddit, for instance, the announcement of a popular AMA (``ask me anything'') session is often accompanied by a ``proof'' post: a recent photograph of the AMA subject (often an entertainer, academic, or social media influencer) shown holding a piece of paper with written confirmation of the date and time of the AMA session and their own Reddit handle \cite{walsh-ama-2023}. Though there are ways to fabricate such posts via image manipulation, the standard for verification is high enough---and evidence of manipulation is often easy enough to detect---that successful spoofs are uncommon, or at least receive little buy-in.

While a full recounting of the various methods for verifying account owner identity is beyond the scope of this work, we note in closing that what matters most is how \textit{effective} an identity confirmation system is. Identity confirmation need not rest on providing official documents or taking photos or videos of a person. Links from authoritative websites or a track record of seemingly authentic conduct are other factors that might play a role, for example. Our recommendation is that when a social media account is verified, regardless of the process, a user on the platform can have high confidence that the account owner is who they claim to be.

\section*{Acknowledgments}
We gratefully acknowledge our anonymous reviewers and shepherd for their insightful feedback. We also thank Andy Guess for reviewing an early version of the survey instrument.

\section*{Availability}

Supplementary information, including the survey instrument, is available at \url{https://github.com/citp/account_verification}.
\printbibliography

\appendix

\section*{Appendix}

\section{Pilot Studies}
\label{sec:pilot}

\textbf{Pilot 1.}
\label{sec:p1}
Our first pilot consisted of 21 questions. At the time of the study, Twitter's subscription blue check program had been live for about three weeks; in addition, gray check marks with ``Official'' labels had recently made an appearance on the platform. Survey respondents selected, from a list of six criteria, those choices that were most likely required for an account to obtain blue checks on Twitter, Facebook, and TikTok; and a gray ``Official'' check on Twitter. 

Their choices indicated uncertainty about the criteria for blue check verification on Twitter---we did not specify if the blue checks in question were paid or ``legacy verified'' (unpaid), only that respondents should evaluate each platform's ``current policies'' for becoming verified---and greater confidence in the verification processes of Facebook and TikTok, whose blue check mark programs are not subscription-based. Interestingly, we noticed that there appeared to be a transferal of trust from Twitter's blue check marks to Twitter's gray ``official'' check marks---some of the verification assurances that respondents associated with Facebook and TikTok's verification marks were likewise strongly associated with Twitter's ``official'' gray checks. 

Following completion of Pilot 1, we added additional questions about respondents' personal experiences with verification. We also added a question to measure digital literacy.

\textbf{Pilot 2.}
By the time we conducted our second pilot study, in January of 2023, Twitter's gray ``Official'' check marks had disappeared, and the platform had introduced gold and gray (minus ``Official'' label) check marks (see~\Cref{table:verif_marks}). Though the stated purpose of the new checks was to differentiate specific types of organizations---gold checks for businesses, and gray checks for government organizations and officials---the new check mark taxonomy was applied unevenly in practice. For a particularly glaring example of the illogic of Twitter's new check mark program, see~\Cref{sec:verif-marks}.

Whereas respondents to Pilot 1 had previously expressed greater trust in gray ``Official'' check marks than in blue check marks on Twitter, all respondents to Pilot 2 expressed confusion about Twitter's unlabeled gray and gold check marks. When asked about the meaning of gray check marks, all respondents wrote ``Do Not Know.'' When asked about the meaning of gold check marks, all respondents but one wrote ``Do Not Know.'' (The sole exception wrote that ``They are notable accounts that were previously verified,'' which is incorrect---gold checks denote verified businesses.) 

Following completion of Pilot 2, we added ``none of the above'' as a multiple choice response to our questions about perceived verification checks. For the same questions, we also added a choice about confirmation of call/text receipt as a verification check. 

\section{User perceptions of verification marks: response taxonomy}
\label{sec:verif_taxon}
In order to establish a taxonomy for inductive coding of survey responses, a researcher read about 290 free responses and noted a handful of recurring responses and terms (e.g., ``factuality,'' ``identity'') and iteratively tagged responses with these terms, expanding this tagging lexicon as new terms appeared. At the end of her initial reading and tagging, the researcher merged any semantically similar tags to produce a final taxonomy. Below is the codebook for our taxonomy of survey respondents' responses to the following question: ``In general, what does a ``verified'' badge on an account mean to you?''

\begin{enumerate}
    \item Account holder identity has been verified: Account is not a bot or spam / fake account. The entity represented by the account is also the entity operating it. 
    \item Account has been vetted by the social media platform: Account has been vetted via a process established by the platform.
    \item Account is influential: Account has a significant number of followers. 
    \item Account belongs to a notable entity: The user represented by the account is a very visible person or organization. In particular, this person or organization was famous in their own right before they joined social media. Verification is necessary in order to deter fake / imitation accounts.  
    \item Account was paid for: The account holder paid for a verification mark.
    \item Account is a credible: Account is a reliable source of information, or has a record of publishing factually correct tweets. 
    \item No meaning: The respondent states that they put little stock in verification marks, ignore them, or feel that these marks have lost all meaning after the introduction of Twitter Blue.  

    \item Uncertain: Respondent is not sure what ``verification'' means.

    \item N/A: No response, or nonsensical response. 
    
\end{enumerate}

\section{Account and content categories}
\label{sec:tweet-codebook}
Below is the codebook from the themes that surfaced during our inductive coding session.

\begin{enumerate}
    \item Business: The official Twitter account representing a large for-profit business.
    \item Small business, or small business owner: An account that primarily promotes a small, independent business.
    \item Organization: The official Twitter account representing a non-commercial, non-government legal entity, like an NGO, non-profit, or trade union.
    \item Politician: Account of an individual in an elected role, or an individual running for office.
    \item Government agency or government official: The Twitter account of a government agency or hired official.
    \item Journalists or news outlets: The Twitter account of a journalist or news outlet, including radio show hosts and other reporters.
    \item Celebrities and influencers: An account belonging an individual, containing self-promotional content. Artists, musicians, content creators, and performers fit in this category. The relative reknown, or follower count, of these accounts does not factor into account classification. 
    \item Personal account: An account belonging to an individual whose tweets primarily contain personal, but not self-promotional, content.
\end{enumerate}
In particular, difficult lines to draw were between ``personal accounts'' and ``influencer'' accounts, especially if the account had a lower number of followers. In addition, it may be difficult to differentiate between ``small business owner'' and ``influencer''. If the account was promoting an individual’s brand, rather than promoting a separate small business, we labelled the account as the latter.

The following emerged as other cohesive trends in the content of account tweets:
\begin{enumerate}
    \item Crypto promoter: If the account content heavily promotes one particular brand of NFT or Cryptocurrency.
    \item Political bent: If the account offers a clear opinion about an elected official.
    \begin{enumerate}
        \item Democrat-leaning: The account tweeted in support of a U.S. Democratic party official, or in opposition of a U.S. Republican party official.
        \item Republican-leaning: The account tweeted in support of a U.S. Republican party official, or in opposition of a U.S. Democratic party official.
        \item Non-U.S. political bent: The account only tweeted opinions on elected officials outside the U.S.
    \end{enumerate}
    \item Retweets or mentions Elon Musk.
    \begin{enumerate}
        \item Mentions are of overwhelmingly positive sentiment, or support.
        \item Mentions are neutral, mixed, or ambiguous. This includes mentions intended to market a product.
        \item Mentions are of overwhelmingly negative sentiment, or criticism.
    \end{enumerate}
\end{enumerate}

\section{News Awareness effects}
\label{sec:attn_check}
We conducted an effect size analysis of respondents who responded correctly to our news-awareness question (\cmark, \textit{n} = 263) and respondents who responded incorrectly (\xmark, \textit{n} = 36). In particular, we compared both groups' responses to our Twitter policy question, which asked respondents to rate the perceived sufficiency of Twitter's current verification policy. 

The Hedges' \textit{g} for both groups was 0.090, indicating negligible effect size; for this reason, we decided to include responses from respondents who failed our attention check question in our final analysis. Some summary statistics for both sample sets follow: 

\begin{table}[!htb]
\small
    \centering
    \begin{tabular}{crrr}
    \toprule
        Group & \# Respondents & Mean & SD \\
        \midrule
        \cmark & 36 & 3.33 & 1.568 \\
        \xmark & 263 & 3.449 & 1.286 \\
    \bottomrule
    \end{tabular}
    \caption{Respondent attention check performance.}
    \label{tab:my_label}
\end{table}

\section{Account types followed}
\label{app:accts_followed}

\begin{figure}[!htb]
\centering
\includegraphics[scale=0.17]{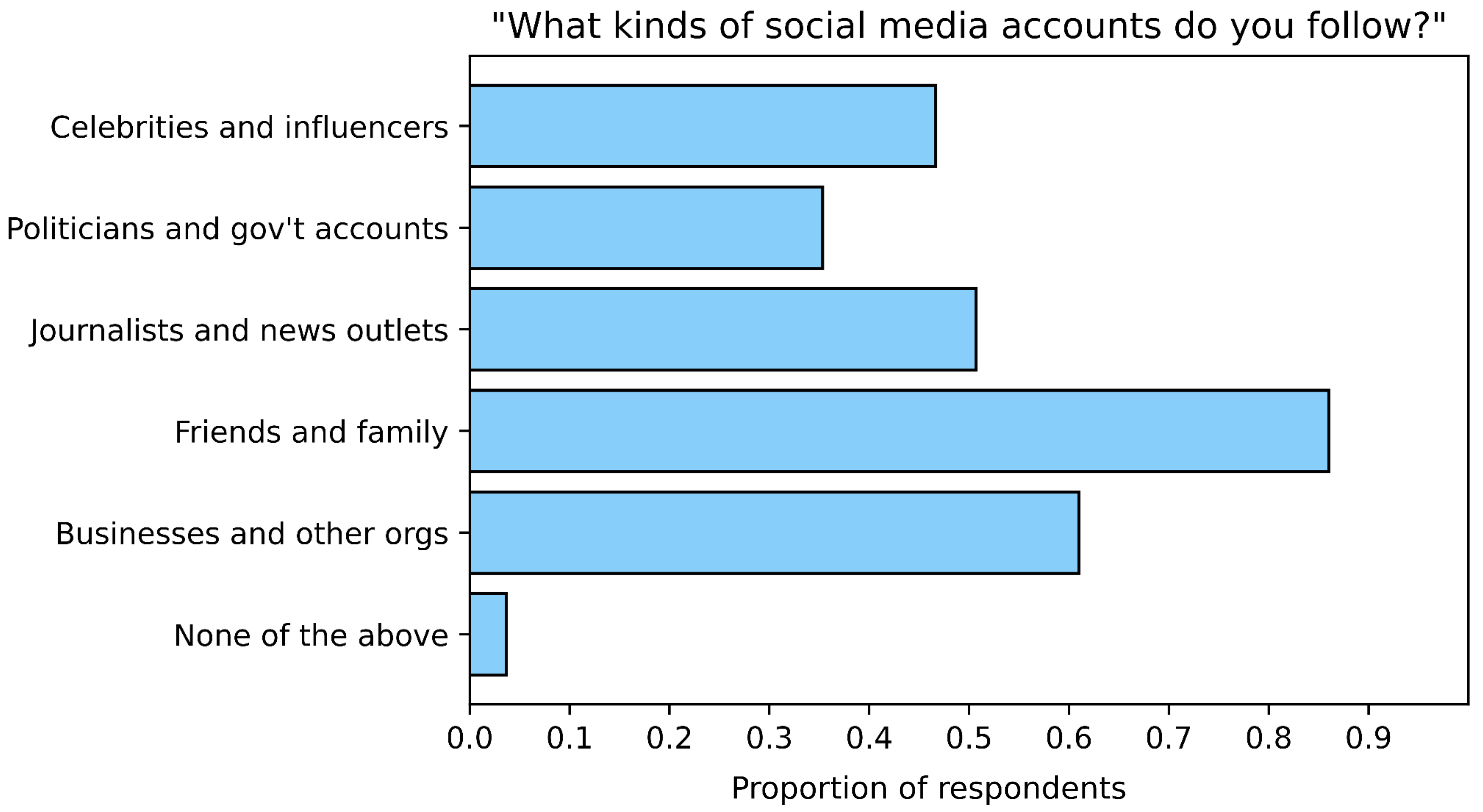}
\caption{In response to a multiple-choice question about social media usage, respondents selected the types of accounts they most commonly follow on social media. 
}
\label{fig:accts_followed}
\end{figure}

\section{Social media verification marks}
\label{sec:verif-marks}

\begin{figure}[!h]
\label{fig:times_checks}
\centering
\includegraphics[scale=0.33]{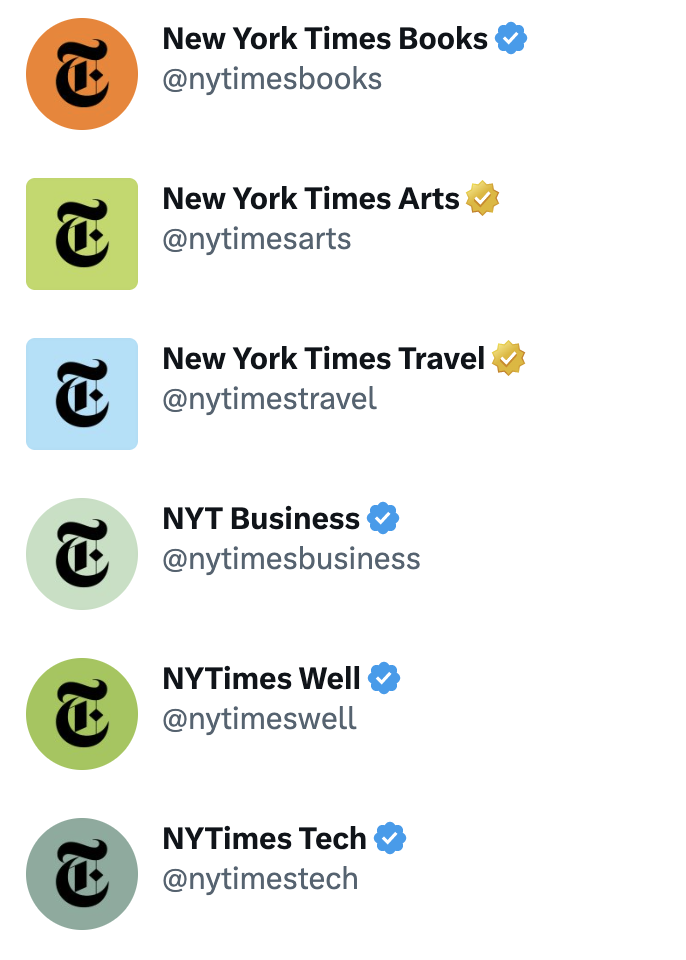}
\caption{Blue and gold checks distributed across New York Times accounts. It is unclear why some sections of the Times received gold checks and others received blue checks. (This snapshot was taken on January 30, 2023.)}
\label{fig:nyt_checks}
\end{figure}

\makeatletter
\setlength{\@fptop}{0pt}
\makeatother

\end{document}